\documentclass{fizik}
\usepackage{amsmath}
\usepackage{amssymb}
\usepackage{psfrag}
\usepackage{graphicx}
\usepackage{psfig}
\usepackage{shortcuts}
\usepackage{longtable}
\usepackage[squaren,thickqspace,thickspace,cdot]{SIunits}

\newcommand{\figtwo}{3in}
\vol{XX}
\pyear{2002}
\received{\today}
\author[TUNCER]{
       {\bf En{\.{\i}}s Tuncer}
        \thanks{{email: {\tt enis.tuncer@physics.org}};
Currently at Department of Physics, University of Potsdam, Potsdam 
14469 Germany.}\\
	{\it Chalmers University of Technology, 
         SE-412\ 96 Gothenburg Sweden}\\ 
	}
\title{How round is round?\\ {\large On accuracy in complex dielectric permittivity calculations:\\ A finite-size 
scaling approach} \thanks{Financed by the ELIS program of the Swedish Foundation for Scientific Research (SSF).}}
\begin{document}
\maketitle

\begin{abstract}
Accuracy in complex dielectric permittivity calculations in binary dielectric mixtures in two-dimensions are reported by taking into account the shape of the inclusion phase. The dielectric permittivity of the mixtures were calculated using the finite element, and the permittivities were estimated by two different procedures. The results were compared with those of analytical models based on mean field approximation and regular arrangement of disks. We have approached the problem emphasizing the finite-size behavior in which regular polygons with $n$ sides were assumed to mimic the disk inclusion phase. It was found that at low concentrations, $<\ 30\ \%$, considering an decagon ($n=10$) cause an error of $<0.1\ \%$ in the effective medium quantities compared with the results obtained using the analytical models. 
\end{abstract}

\section{Introduction}
To predict and to better design (tailor) composite materials for electrical applications, such as composite insulators~\cite{TuncerLic,TuncerPhD} and elecromagetic shields~\cite{Pier6}, {\em etc}, have been a challenge for both theoretical and practical importance~\cite{SihvolaBook}. In early days of electromagnetics theory, effective electrical properties, {\em i.e.}, conductivity, $\sigma$, and permittivity, $\epsilon$, of systems composed of two phases have been calculated analytically by effective mean field approaches~\cite{Maxwell_Garnett,wiener,Wagner1914,Sillars1937,Steeman90}  ({\sc ema}) and regular arrangement of disks~\cite{Rayleigh,Emets98b} ({\sc rad}). And nowadays, computer simulations have become as an alternative way of doing science closer to experiment than theory but complementory to both. Moreover, with the help of new computation techniques to solve partial differential equations, numerical simulations of more complex systems, such as systems with several components with arbitrary shapes, can be considered and desired properties can be calculated.  However, like experiments, computer simulations produce data rather than theories and should be judged on the quality of those data. Accodingly, the reliability of the applied technique and accuracy of the obtained results must be checked and be verified either by analytical solutions or by performing the same calculations with a different technique or tool~\cite{Experiment}. The latter procedure can be, for example, applications of the finite element and the finite difference methods to the same problem. The former one, to verify with an analyical solution, is not, on the other hand, an easy task in most of the cases, in which either there are no analytical solutions or approximations are considered in the analytical solutions. Moreover, when the numerical results are taken into consideration, there are plausible sources of errors which originate from the model. The numerical errors can be eliminated by changing the order of the applied model or the discritization method used. 

In this paper, the accuracy in the numerical calculations of effective electrical properties of a binary dielectric mixture is reported by taking into account the finite-size scaling. A field simulation software, based on the finite element method ({\sc fem}), has been used, and the effective properties has been calculated by two different ways. The results were compared with those of two anaytical solutions, which are based on {\sc ema} and  {\sc rad}.

\section{Electrical properties of binary mixtures}
\begin{figure*}[tp]
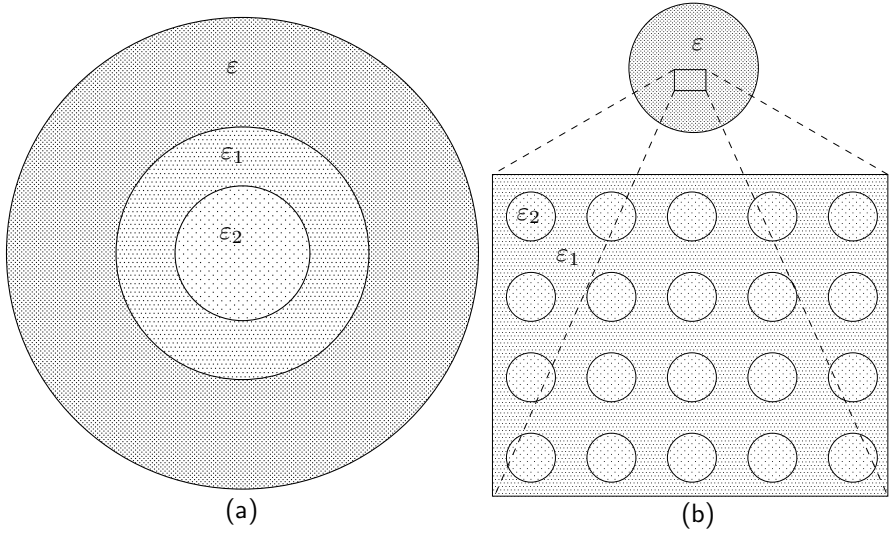

  \centering{\psfragscanon
  \psfrag{1}[][]{\bf $\varepsilon_2$}
  \psfrag{2}[][]{\bf $\varepsilon_1$}
  \psfrag{3}[][]{\bf $\varepsilon$}
  \psfrag{(a)}[][]{{\sf (a)}}
  \psfrag{(b)}[][]{{\sf (b)}}
  \includegraphics[width=2.5in]{cyl.eps}
  \includegraphics[width=2.1in]{cyl2.eps}
  \psfragscanoff}
  \caption{(a) {\sc ema} and (b) {\sc rad} approaches to dielectric mixtures.\label{fig:Cyl}}
\end{figure*}

Electrical properties of materials can be described by the dependence of either the complex dielectric permittivity, $\varepsilon(\omega)$, or complex ({\sc ac}) conductivity, $\varsigma(\omega)$, on the frequency, $\omega$ and on the external variables such as temperature, pressure, humidity {\em etc.} Both of these quantities can be expressed in terms of one and other. Therefore, a general complex dielectric response of materials can be describe with the complex dielectric susceptibility, $\chi(\omega)$,
\begin{eqnarray}
  \label{eq:epssig}
  \varepsilon(\omega)&=&\epsilon+\chi'(\omega)-\imath\chi''(\omega)+\frac{\sigma}{\imath\epsilon_0\omega}\\
  \varsigma(\omega)&=&\imath\epsilon_0\omega\varepsilon(\omega)
\end{eqnarray}
where $\imath=\sqrt{-1}$ and $\epsilon_0$ is the dielectric permittivity of free space, $1/36\pi\ \nano\farad\per\meter$. $\epsilon$ and $\sigma$ are the dielectric constant at optical frequencies and ohmic conductivity of the material, respectively. Moreover, the simplest form of dielectric relaxation, $\chi(\omega)$, is oberved for dilute solutions (mixtures) and ferroelectric materials~\cite{Jonscher1983}, and is expressed in Debye form~\cite{Debye1945}
\begin{eqnarray}
  \label{eq:debye}
  \chi(\omega)&=&\chi(0)[1+\imath\omega\tau]^{-1}
\end{eqnarray}
where $\chi(0)$ and $\tau$ are the dielectric strength and relaxation time (inverse relaxation rate) of the polarization. At frequencies much higher and lower than inverse relaxation time, $\tau^{-1}$, there are only three material quantities that explicate electrical properties, $\epsilon$, $\chi(0)$ and $\sigma$,
\begin{displaymath}
  \genfrac{}{}{0pt}{}{\varepsilon(\omega)=\epsilon+\displaystyle\frac{\sigma}{\imath\epsilon_0\omega}}{\varsigma(\omega)=\sigma+\imath\epsilon_0\epsilon\omega} \quad  \text{for}  \quad \omega\gg\tau^{-1}
\end{displaymath}
\begin{displaymath}
   \genfrac{}{}{0pt}{}{\varepsilon(\omega)=\epsilon+\chi(0)+\displaystyle\frac{\sigma}{\imath\epsilon_0\omega}}{\varsigma(\omega)=\sigma+\imath\epsilon_0[\epsilon+\chi(0)]\omega} \quad  \text{for} \quad \omega\ll\tau^{-1}
\end{displaymath}

When the analytical solutions of binary mixtures in two-dimensions are considered, the {\sc ema} approach supposes two concentric dielectric disks with dielectric permittivities $\varepsilon_1$ and $\varepsilon_2$ such that they are embedded inside the effective medium with dielectric permittivity, $\varepsilon$,  as presented in Fig.~\ref{fig:Cyl}a. In the two-dimensional {\sc rad} approach, on the other hand, the composite medium is assumed to be composed of monodispersed inclusions (disks) at square lattice sides, as illustrated in Fig.~\ref{fig:Cyl}b. The effective complex dielectric permittivities calculated by {\sc ema}~\cite{Maxwell_Garnett,wiener,Steeman90} and {\sc rad}~\cite{Rayleigh,Emets98b}, then,  yield
\begin{eqnarray}
  \label{eq:ema}
 \varepsilon_{\rm EMA}=\frac{\varepsilon_{2}\,\varepsilon_{1} + {\varepsilon_{1}}^2 +  \varepsilon_{2}\,\varepsilon_{1}\,q - {\varepsilon_{1}}^2\,q}{ \varepsilon_{2} + \varepsilon_{1} - \varepsilon_{2}\,q + \varepsilon_{1}\,q}
\end{eqnarray}

\begin{eqnarray}
  \label{eq:emets}
        \varepsilon_{\rm RASS}&=&\varepsilon_1\frac{\pi-\pi\Lambda q+4\Lambda^2 q(A+B)}{\pi+\pi\Lambda q+4\Lambda^2 q(A+B)}\nonumber \\
 \Lambda&=&\frac{\varepsilon_1-\varepsilon_2}{\varepsilon_1+\varepsilon_2}
\end{eqnarray}
where, $q$ is the concentration of the inclusion phases, which are denoted by subscript (2), ($0\le q \le 1$). The parameters $A$ and $B$ are functions of the radius of the inclusion phase, $r\equiv \sqrt{q/\pi}$,
\begin{eqnarray} \label{A}
 A&=&2r^3\sum_{m=1}^{\infty} \frac{1}{r^4-16m^4}
    +\sum_{n=1}^{\infty}\sum_{m=1}^{\infty} \biggl[
    \frac{r-2m}{(r-2m)^2-4n^2} \nonumber \\ 
    &&+ \frac{r+2m}{(r+2m)^2-4n^2}+\frac{r-2m+1}{(r-2m+1)^2-(2n-1)^2}\nonumber \\
    &&+\frac{r+2m-1}{(r+2m-1)^2-(2n-1)^2}    \biggr] 
\end{eqnarray}
\begin{eqnarray}    \label{B}
 B&=&2r^3\sum_{m=1}^{\infty} \frac{1}{r^4-(2m-1)^4}\nonumber \\
    &&+\sum_{n=1}^{\infty}\sum_{m=1}^{\infty} \biggl[\frac{r-2m}{(r-2m)^2-(n-1)^2}\nonumber \\
  &&+\frac{r+2m}{(r+2m)^2-(n-1)^2}\frac{r-2m+1}{(r-2m+1)^2-4n^2}\nonumber \\
 &&+\frac{r+2m-1}{(r+2m-1)^2-4n^2} \biggr]
\end{eqnarray}
$A$ and $B$ values converge quickly a constant value for $n=m\ge10$. 

\section{Numerical calculations}

Analytical calculations of electromagnetic problems using Maxwell's equations, are limited to geometrical constraints. For some simple geometries with a small number of materials (regions) and symmetries, analytical solutions can be found~\cite{Weber,eme96,Emets98b,Jackson1975,Bottcher}. The analytical solutions are obtained using methods of images~\cite{Elliot,Eyges}, orthogonal functions (Green functions)~\cite{Steiner} and complex variable techniques~(conformal mapping)~\cite{Ramo,Eyges,Weber}. The conformal mapping can only be applied to two-dimensional problems in which the third spatial axis is neglected. 
For more complex geometries and non-homogeneous regions composed of several materials, numerical solutions of partial differential equations and of integral equations have been developed~\cite{Booton} {\em e.g.}, the finite difference method, the finite element method, the method of moments and the boundary element method. 

\begin{figure}[tp]
  \begin{center}
    \psfragscanon
    \psfrag{A}[][]{A}
    \psfrag{B}[][]{B}
    \psfrag{C}[][]{C}
    \psfrag{D}[][]{D}
    \psfrag{P}[][]{P}
    \resizebox{\figtwo}{!}{\includegraphics{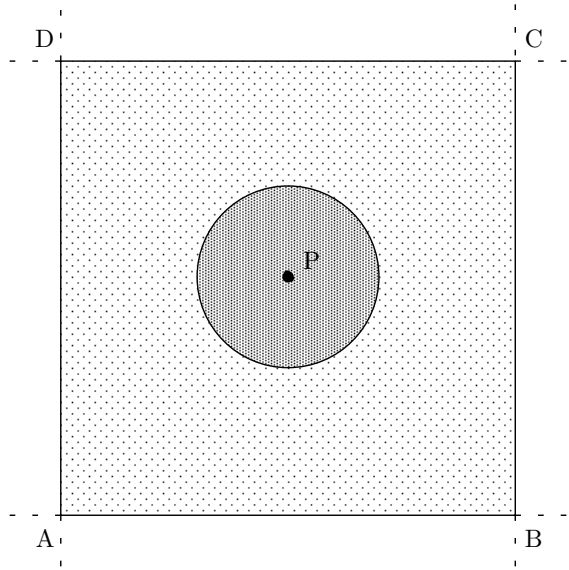}}
  \end{center}
  \caption{\label{fig:unit}The unit cell of square lattice with corners at $ABCD$ used in the calculations. Dark region is the disk inclusion with $\varepsilon_2$ and $\sigma_2$ and the lighter region is the matrix media with  $\varepsilon_1$ and $\sigma_1$.}
\end{figure} 

Numerical solutions of electrostatic problems within a non-conducting medium are based on solving Poisson's equation 
\begin{equation}
  \label{eq:Poisson}
  {\mathbf \nabla} \cdot (\epsilon\epsilon_0 {\mathbf \nabla} \phi) = -\rho
\end{equation}
where $\phi$, and  $\rho$ denote the electrical potential and the total charge in the considered region, respectively. Moreover, if the medium is conductive where no free charges and sources of charges are allowed, then, the solution is given by
\begin{equation}
  \label{eq:Laplace}
  {\mathbf \nabla} \cdot (\sigma {\mathbf \nabla} \phi) = 0
\end{equation}
When the medium is a mixture of these two cases (lossy dielectric), it consists of dielectric and conductive components. The solution, then, becomes time dependent and is given by a complex electric potential in the region with the coupling of Eqs.\ (\ref{eq:Poisson}) and (\ref{eq:Laplace}), which is also known as the continuity equation. 
\begin{equation}
    {\mathbf \nabla} \cdot \left(\sigma {\mathbf \nabla} \phi \right) + {\mathbf \nabla} \cdot \left[\frac{\partial}{\partial t} \left(\epsilon\epsilon_0 {\mathbf \nabla} \phi \right)\right] = 0
    \label{eq:contin.1}
\end{equation}
or equivalently in Fourier-space with frequency dependent properties,
\begin{equation}
    {\mathbf \nabla} \cdot \left\{\left[ \imath \epsilon_0\varepsilon(\omega)\omega \right] {\mathbf \nabla} \phi \right\} = 0
    \label{eq:contin.2}
\end{equation}
where no free charges are allowed in the region, due to conductivity of the medium (lossy dielectric). Note that $\varepsilon(\omega)$ in Eq.~(\ref{eq:contin.2}) is given in Eq.~(\ref{eq:epssig}). 

In this work, we have used a field calculation software, `\textsc{Ace}',~\cite{AceManual} based on the {\sc fem}. A square lattice unit-cell with a hard-disk inclusion was assumed, as shown in Fig. \ref{fig:unit}. The boundary conditions were chosen as follows; along line $[\text{AB}]$ was the ground potential level, $V=0\ \volt$, along line $[\text{CD}]$ it was $1\ \volt$, and the lines $[\text{AD}]$ and $[\text{BC}]$ were the symmetry lines (axes of reflection). The calculations were performed under steady-state periodic conditions. The region of interest is meshed using an triangular meshing technique in which we have limited number of triangles by $\sim8000$ elements. The minimum triangle size was selected using the size of the considered inclusion geometry in the meshing procedure. Moreover, the quadratic shape function was used to solve in the {\sc fem}.

\begin{table}[tp]
  \caption{Composite electric properties obtained from the analytical formulae.\label{tab:analytic}}
\begin{center}
    \begin{tabular}{lrrrr}
\br   
   Model &$q$& $\varepsilon_{\mega}$ & $\varepsilon_{\micro}$ & $\sigma/\varepsilon_0\ [\siemens\per\farad]$ \\
\mr
      {\sc ema}  &0.1& 2.28571429 & 2.42552966 & 0.13749080 \\
      {\sc rad} &0.1& 2.28586956 & 2.42602610 & 0.13752067 \\
\mr
      {\sc ema}  &0.2& 2.61538462 & 2.95231624 & 0.16802403 \\
      {\sc rad} &0.2& 2.61743641 & 2.95942150 & 0.16845390 \\
\mr
      {\sc ema}  &0.3& 3.00000000 & 3.62134985 & 0.20703632 \\
      {\sc rad} &0.3& 3.01008959 & 3.66005320 & 0.20939659 \\
\br
    \end{tabular}
\end{center}
\end{table}

The complex permittivity of a heterogeneous medium can be calculated in several ways, {\em e.g.}, (i) by using the total current density, $j$, and the phase difference, $\theta$~\cite{vonHippel,Scaife,Tuncer2001a,Tuncer2001b}, (ii) Gauss' law and losses~\cite{Scaife,Sareni1996,Tuncer1998b} and lastly (iii) by using the average values of dielectric displacement $\langle{\mathbf{D}}\rangle$ and electric field $\langle{\mathbf{E}}\rangle$~\cite{Scaife,LL}. We have used  the first two ways (i \& ii) to calculate the complex dielectric permittivities, $\varepsilon(\omega)$ of the structures considered. The phase parameters, $\varepsilon$ and $\sigma$, were frequency and voltage independent. The values of $\varepsilon$ and $\sigma$ are chosen such that the interfacial polarization observed has a relaxation time, $\tau$, around $1\ \second$. This is achieved when the matrix phase has $\varepsilon_1\equiv\epsilon_1=2$ and $\sigma_1=1\ \pico\siemens\per\meter$, and the inclusion phase has $\varepsilon_2\equiv\epsilon_2=10$ and $\sigma_2=100\ \pico\siemens\per\meter$. The normalized conductivity values, $\sigma/\epsilon_0$, are $\sigma_1/\epsilon_0=0.113\ \siemens\per\farad$ and $\sigma_2/\epsilon_0=11.3\ \siemens\per\farad$, respectively. We have focused on two frequencies, $2\pi \micro\hertz$ and $2\pi \mega\hertz$ which are,  respectively, used as subscripts ($\micro$) and ($\mega$) for the appropriate dielectric permittivity values. At these frequencies, the influence of interfacial polarization was negligible, $|\log\omega|\gg\tau^{-1}$, and the composite medium can be expressed with three parameters, $\varepsilon_{\mega}$, $\varepsilon_{\micro}~[\equiv\varepsilon_{\mega}+\chi(0)]$ and $\sigma/\epsilon_0$.
\begin{figure}[t]
 	\begin{center}
	{\includegraphics[width=3.8in]{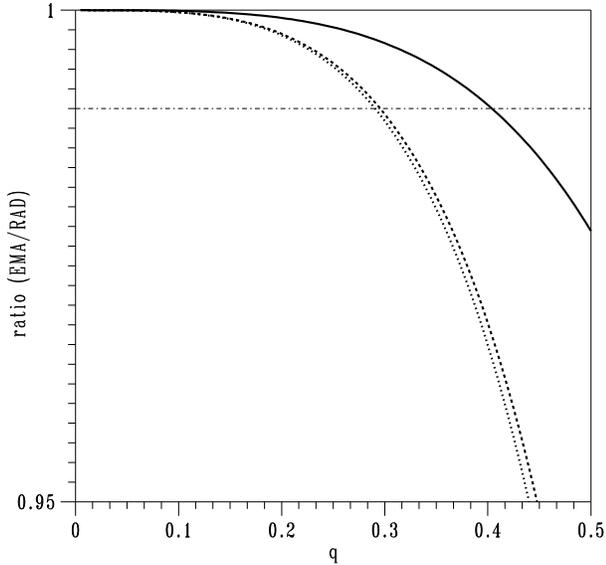}}
	\end{center}
  \caption{\label{fig:ratio}Comparison of the two analytical models. The solid ($\full$), dashed ($\dashed$) and dotted ($\dotted$) lines represent the ratios of the high frequency dielectric permittivity, $\varepsilon_{\mega}$, low frequency permittivity, $\varepsilon_{\micro}$, and the conductivity, $\sigma/\epsilon_0$, values, respectively. The region above the thin chain ($\chain$) line marks the $1\ \%$.} 
\end{figure}

\section{Results and discussion}
\begin{figure}[b]
  \begin{center}
    \psfragscanon
    \psfrag{r3}[][]{$r_3$}
    \psfrag{r4}[][]{$r_4$}
    \psfrag{r5}[][]{$r_5$}
    \psfrag{r0}[][]{$r_{\infty}$}
    \psfrag{n3}[][]{$n=3$}
    \psfrag{n4}[][]{$n=4$}
    \psfrag{n5}[][]{$n=5$}
    \psfrag{n0}[][]{$n={\infty}$}
    \rotatebox{0}{\includegraphics[height=3in]{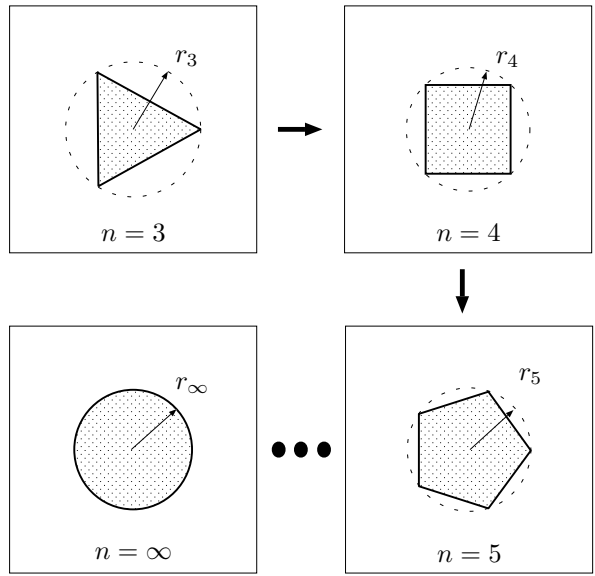}}
    \psfragscanoff
  \end{center}
  \caption{Finite-size scaling of the inclusion shapes (polygons).\label{fig:Shapes}}
\end{figure}
The concentration dependence of the electrical properties calculated using the analytical formulae of Eqs.~(\ref{eq:ema}) and (\ref{eq:emets}) differ at some concentration level. This is ilustrated in Fig.~\ref{fig:ratio} by the ratio of resulting effective parameters. For concentration values higher than $30\ \%$ ($q>0.3$), the effective medium quantities obtained from the two analytical approaches start to differ. The behavior of ratios for the dielectric permittivity at low frequencies, $\varepsilon_{\micro}$ and the conductivity, $\sigma$, were similar and the change with respect to concentration, $q$, was steeper compared to the ratio of the dielectric permittivity at low frequencies, $\varepsilon_{\mega}$. The difference between the models is due to the approximations and simplifications considered in the geometries. In {\sc rad} assumptions there are neighboring inclusions whose charge distributions (polarization) influence the polarization of the individual inclusions, however, in {\sc ema} the polarization of the inclusion is only due to the interface between the inclusion and the matrix phases. Accordingly, in the numerical simulations, three concentration levels were selected, $q=\{0.1,0.2,0.3\}$, and the $\varepsilon_{\mega}$-, $\varepsilon_{\micro}$- and $\sigma/\epsilon_0$-values were calculated. The same quantities obtained from {\sc ema} and {\sc rad} are listed in Table~\ref{tab:analytic} for comparison. The differences in the $\varepsilon_{\mega}$ and $\sigma/\epsilon_0$ values calculated from the analytical models are in the fourth, third and second number after the decimal point for $q={0.1,0.2,0.3}$, respectively. However, the change in $\varepsilon_{\micro}$ is larger for both models at the same concentrations.

In the simulations, we have assumed that the inclusion phase was a two-dimensional object, a {\em regular} polygon with $n$ sides, as displayed in Fig.~\ref{fig:Shapes}. The polygons were generated using a circle with radius, $r$, and a contraint on the area of the polygons, $q$. Then, the radius, $r$, as a function of $n$ and $q$ is expressed as,
\begin{equation}
  \label{eq:radius}
  r_n=\sqrt{\genfrac{}{}{1pt}{}{q}{n \sin(\pi/n)\cos(\pi/n)}}
\end{equation}
The denominator inside the square root approaches $\pi$ as $n\rightarrow\infty$. The size of radius was also used for the meshing procedure of the computation domain where $r_n/15$ was the size of the minimum triangle. Furthermore, this approach leads to the finite-size scaling considerations~\cite{Plischke1994,finite-size} in which as $n\rightarrow\infty$, the inclusion phase is a perfect disk. As mentioned previously, two different methods were used to calculate the electrical quantities of the binary mixture, and Table~\ref{tab:numeric1} presents the results. The first remark was that the difference between the obtained $\varepsilon_{\micro}$-values using two different approaches. The discripancy between $\varepsilon_{\micro}^{\text{a}}$ and $\varepsilon_{\micro}^{\text{b}}$ increased as the concentration level, $q$, was increased. Moreover, when the values were compared top those of Table~\ref{tab:analytic}, except $\varepsilon_{\micro}^{\text{b}}$, the other calculated quantities had good agreement with the analytical models. 
\begin{figure}[t]
\centering{\includegraphics[width=3in,angle=-90]{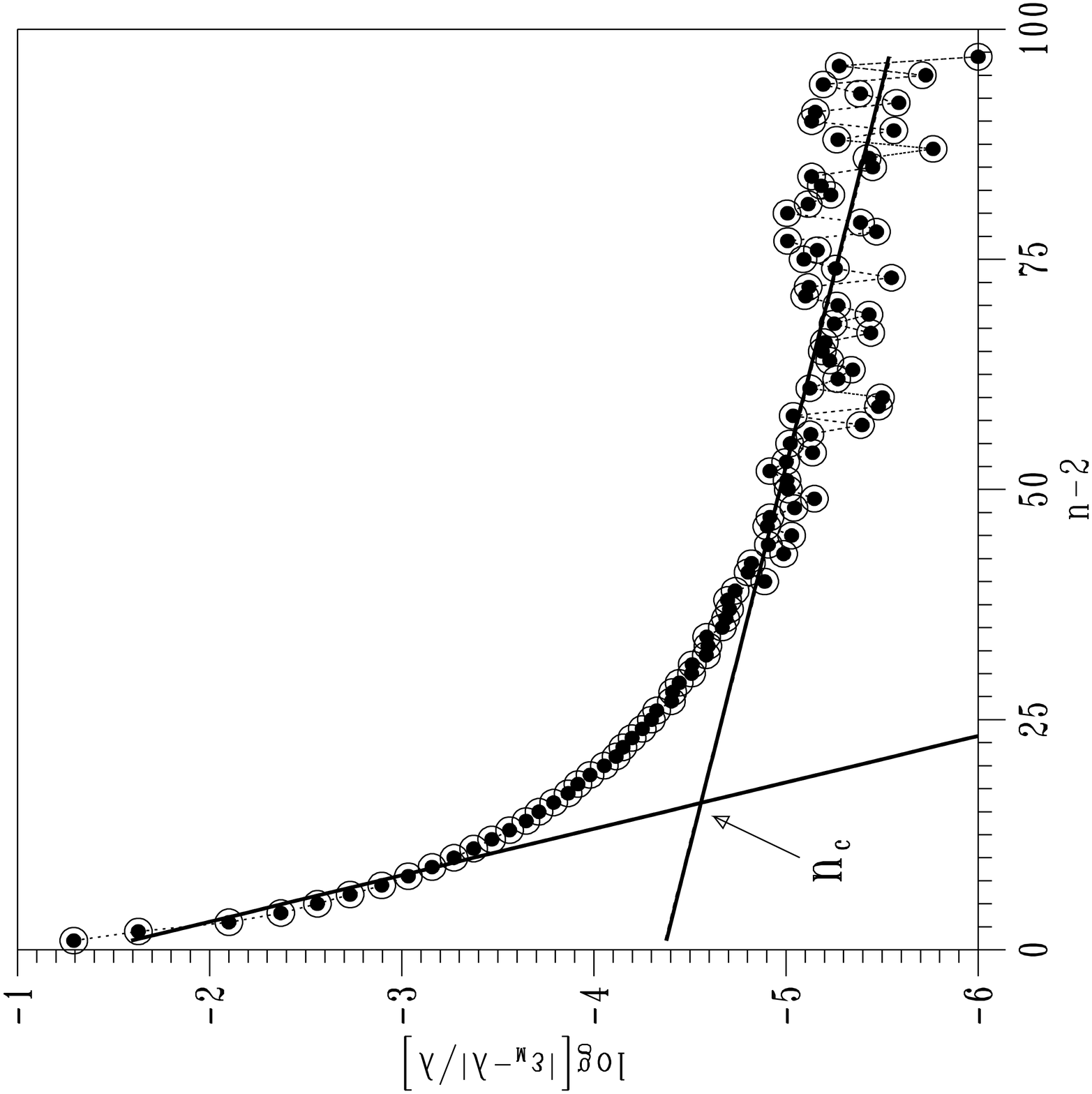}}
\caption{Normalized dependence of incrimental high frequency dielectric permittivity, $|\varepsilon_{\mega}-\lambda|/\lambda$ on number of regular polygon sides, $n$. The symbols open ($\bigcirc$) and filled ($\bullet$) indicate the solutions obtained using the current density and phase shift between the applied voltage and current and using Gauss' law and the total losses in the medium, respectively. The solid lines ($\full$) represents the fitted curves. $n_c$ is the critical side number for regular polygons. \label{fig:fittings4}}
\end{figure}

\begin{table*}[tp]
  \begin{center}
  \caption{Electrical parameters calculated by the {\sc fem}.\label{tab:numeric1}\label{tab:numeric2} \label{tab:numeric3}}
\begin{tabular}{rrrrrrr}
\br
$n$ & $\varepsilon_{\mega}^{\text a}$ & $\varepsilon_{\micro}^{\text a}$ & $\sigma/\varepsilon_0^{\text a}\ [\siemens\per\farad]$ & $\varepsilon_{\mega}^{\text b}$ & $\varepsilon_{\micro}^{\text b}$ & $\sigma/\varepsilon_0^{\text b}\ [\siemens\per\farad]$ \\
\mr
$q=0.1$&&&&&& \\ 
\mr
 3 & 2.31830984 & 2.55431772 & 0.14540117 & 2.31831000 & 2.57482200 & 0.14462477 \\ 
 4 & 2.29814032 & 2.46885006 & 0.14011433 & 2.29814046 & 2.48120081 & 0.13990578 \\ 
 5 & 2.29154397 & 2.44497599 & 0.13866220 & 2.29154405 & 2.45548560 & 0.13852250 \\ 
 6 & 2.28903383 & 2.43636313 & 0.13814092 & 2.28903405 & 2.44625458 & 0.13804715 \\ 
 7 & 2.28780142 & 2.43222046 & 0.13789067 & 2.28780167 & 2.44182314 & 0.13782876 \\ 
 8 & 2.28712945 & 2.43000429 & 0.13775702 & 2.28712965 & 2.43945636 & 0.13769620 \\ 
 9 & 2.28672438 & 2.42867095 & 0.13767662 & 2.28672448 & 2.43803276 & 0.13762992 \\ 
10 & 2.28646432 & 2.42782513 & 0.13762567 & 2.28646448 & 2.43713033 & 0.13759796 \\ 
11 & 2.28629434 & 2.42727218 & 0.13759236 & 2.28629442 & 2.43654060 & 0.13756319 \\ 
12 & 2.28617467 & 2.42688644 & 0.13756913 & 2.28617471 & 2.43612935 & 0.13754747 \\ 
14 & 2.28602947 & 2.42641794 & 0.13754093 & 2.28602973 & 2.43562974 & 0.13752268 \\ 
16 & 2.28594550 & 2.42615013 & 0.13752482 & 2.28594567 & 2.43534447 & 0.13750994 \\ 
18 & 2.28589649 & 2.42599233 & 0.13751533 & 2.28589650 & 2.43517637 & 0.13750239 \\ 
99 & 2.28577330 & 2.42560592 & 0.13749210 & 2.28577345 & 2.43476492 & 0.13748560 \\ 
$\infty$ & 2.28576594 & 2.42558544 & 0.13749087 & 2.28576605 & 2.43474329 & 0.13747696 \\
\mr
$q=0.2$&&&&&& \\ 
\mr
3 & 2.69416747 & 3.30221747 & 0.18974833 & 2.69416760 & 3.36013939 & 0.18911785 \\ 
4 & 2.64801022 & 3.07703286 & 0.17563592 & 2.64801032 & 3.11023106 & 0.17541128 \\ 
5 & 2.62892034 & 3.00133543 & 0.17099827 & 2.62892033 & 3.02810579 & 0.17088012 \\ 
6 & 2.62304978 & 2.97945868 & 0.16966607 & 2.62304982 & 3.00451441 & 0.16958761 \\ 
7 & 2.62025295 & 2.96934543 & 0.16905189 & 2.62025298 & 2.99363860 & 0.16899605 \\ 
8 & 2.61869852 & 2.96379171 & 0.16871499 & 2.61869871 & 2.98767247 & 0.16866931 \\ 
9 & 2.61776224 & 2.96046863 & 0.16851354 & 2.61776238 & 2.98410506 & 0.16847596 \\ 
10 & 2.61716591 & 2.95837257 & 0.16838656 & 2.61716598 & 2.98185662 & 0.16836172 \\ 
11 & 2.61677916 & 2.95701332 & 0.16830424 & 2.61677923 & 2.98039874 & 0.16827771 \\ 
12 & 2.61650596 & 2.95606556 & 0.16824686 & 2.61650611 & 2.97938287 & 0.16822349 \\ 
14 & 2.61616711 & 2.95488967 & 0.16817569 & 2.61616735 & 2.97812228 & 0.16815940 \\ 
16 & 2.61598326 & 2.95424269 & 0.16813652 & 2.61598338 & 2.97742871 & 0.16812236 \\ 
99 & 2.61560667 & 2.95293358 & 0.16805731 & 2.61560681 & 2.97602608 & 0.16806089 \\ 
$\infty$ & 2.61561521 & 2.95297927 & 0.16806011 & 2.61561540 & 2.97607565 & 0.16805491\\
\mr
$q=0.3$&&&&&& \\ 
\mr
3 & 3.15432863 & 4.44992747 & 0.25956453 & 3.15432877 & 4.59647254 & 0.25812644 \\ 
4 & 3.07199643 & 3.93508366 & 0.22640982 & 3.07199661 & 4.00935583 & 0.22583378 \\ 
5 & 3.02527912 & 3.72286805 & 0.21324951 & 3.02527918 & 3.77630766 & 0.21302790 \\ 
6 & 3.01422774 & 3.67689028 & 0.21042426 & 3.01422781 & 3.72627694 & 0.21031765 \\ 
7 & 3.00968689 & 3.65880191 & 0.20931774 & 3.00968704 & 3.70668244 & 0.20923851 \\ 
8 & 3.00700955 & 3.64824288 & 0.20867221 & 3.00700972 & 3.69525097 & 0.20859681 \\ 
9 & 3.00525073 & 3.64134674 & 0.20825088 & 3.00525070 & 3.68779013 & 0.20820236 \\ 
10 & 3.00420698 & 3.63728448 & 0.20800287 & 3.00420720 & 3.68339788 & 0.20795785 \\ 
11 & 3.00352151 & 3.63462285 & 0.20784042 & 3.00352166 & 3.68052130 & 0.20781141 \\ 
12 & 3.00303949 & 3.63276735 & 0.20772723 & 3.00303968 & 3.67851699 & 0.20769857 \\ 
14 & 3.00245464 & 3.63050408 & 0.20758916 & 3.00245485 & 3.67607184 & 0.20756988 \\ 
16 & 3.00211061 & 3.62921448 & 0.20751059 & 3.00211061 & 3.67468077 & 0.20749301 \\ 
99 & 3.00143716 & 3.62665177 & 0.20735438 & 3.00143741 & 3.67191430 & 0.20734844 \\ 
$\infty$ & 3.00147224 & 3.62677466 & 0.20736184 & 3.00147232 & 3.67204654 & 0.20735887 \\
\br
\end{tabular}\\%
{\small $^{\text a}$ Calculated from the current and phase shift between the applied voltage and current.\\
$^{\text b}$ Calculated using Gauss' law and the total losses in the medium.}    
  \end{center}
\end{table*}

Obtained quantities, $\varepsilon_{\mega}$-, $\varepsilon_{\micro}$- and $\sigma/\epsilon_0$, can be described by a trivial relation, which considers the finite-size behavior,
\begin{equation}
  \label{eq:finitesize}
  f(n,q)-\lambda\approx a_1(n-2)^{\alpha_1(q)}+a_2(n-2)^{\alpha_2(q)}
\end{equation}
where $\lambda=f(\infty,q)$. We have also scaled the above equation with $n-2$ since the calculations were performed in two-dimensions. The $\lambda$-values are presented in Table~\ref{tab:numeric1} as $n\rightarrow\infty$, except $\varepsilon_{\micro}^{\text{b}}$ all others were close to the data in Table~\ref{tab:analytic}. In Fig.~\ref{fig:fittings4}, an example of the finite-size behavior is shown. A critical number of sides, $n_c$, is defined, such that over this value, $n>n_c$, the effective properties of medium with regular polygons as inclusions are approximately similar to those of a medium with disk shape inclusions. The analysis showed that $n_c$ is approximately $15$ regardless the concentration levels considered, $q\le0.3$, and the error in the calculations is $<0.01\ \%$ for $n_c>15$, as displayed in Fig.~\ref{fig:fittings4}.  In fact, even an decagon ($n=10$) can imitate a disk in which the error in the calculated electrical quantities is less than $<0.1\ \%$. Furthermore, Eq.~(\ref{eq:finitesize}) is divided in two,

\begin{eqnarray}
f(n,q)-\lambda\propto\left\{ \begin{array}{ll}
(n-2)^{\alpha_1(q)} & n<15\\
(n-2)^{\alpha_2(q)} & n>15 \end{array} \right.
\end{eqnarray}
In Table~\ref{table:fits}, the parameters of this behavior, Eq.~(\ref{eq:finitesize}), from a curve-fitting procedure are presented. All calculated quantities had similar behavior as in the Fig.~\ref{fig:fittings4}. It was clear to us that the exponents were concentration independent, $\alpha\neq F(q)$. Moreover, although $\alpha_1$ was constant for all considered quantities, $\alpha_2$ was dependent on quantities. Finally, there was no trivial relation between concentration and obtained $a$ values.
\begin{table*}[tp]
  \caption{Finite-size behavior modeling parameters.  \label{table:fits}}
\begin{center}
  \begin{tabular}[ ]{lrrrrrrrrr}
\br
&$q$&$\log a_1^{\text{a}}$ & $\alpha_1^{\text{a}}$ & $\log a_2^{\text{a}}$ & $\alpha_2^{\text{a}}$ & $\log a_1^{\text{b}}$ & $\alpha_1^{\text{b}}$ & $\log a_2^{\text{b}}$ & $\alpha_2^{\text{b}}$ \\
\mr
$\varepsilon_{\mega}$   & 0.1 & -1.500 & -0.215 & -4.174 & -0.017 & -1.500 & -0.215 & -4.172 & -0.017 \\
$\varepsilon_{\micro}$  & 0.1 & -0.930 & -0.223 & -3.530 & -0.022 & -0.895 & -0.224 & -3.487 & -0.023 \\
$\sigma/\epsilon_{0}$ & 0.1 & -2.145 & -0.224 & -5.077 & -0.010 & -2.178 & -0.230 & -5.014 & -0.003 \\
\mr
$\varepsilon_{\mega}$   & 0.2 & -1.104 & -0.220 & -3.994 & -0.015 & -1.104 & -0.220 & -3.990 & -0.015 \\
$\varepsilon_{\micro}$  & 0.2 & -0.492 & -0.230 & -3.240 & -0.022 & -0.455 & -0.231 & -3.195 & -0.022 \\
$\sigma/\epsilon_{0}$ & 0.2 & -2.202 & -0.230 & -5.253 & -0.008 & -2.210 & -0.233 & -5.330 & -0.005 \\
\mr
$\varepsilon_{\mega}$   & 0.3 & -0.806 & -0.227 & -3.724 & -0.015 & -0.806 & -0.227 & -3.722 & -0.015 \\
$\varepsilon_{\micro}$  & 0.3 & -0.127 & -0.241 & -2.970 & -0.021 & -0.083 & -0.242 & -2.926 & -0.021 \\
$\sigma/\epsilon_{0}$   & 0.3 & -1.829 & -0.242 & -4.951 & -0.011 & -1.840 & -0.243 & -4.993 & -0.009 \\
\br
  \end{tabular}\\
{\small $^{\text a}$ Obtained for values calculated by using the current and phase shift between the applied voltage and current.\\
$^{\text b}$ Obtained for values calculated by using Gauss' law and the total losses in the medium.}
\end{center}
\end{table*}

\section{Conclusions}

Calculations of effective electrical properties of binary dielectric mixtures were used to evaluate number of sides of a regular polygon in order the considered polygon to imitate a disk in computer calculations. We assumed a meduim with inclusions as regular polygons with $n$ sides, and calculated the electrical quantities, {\em i.e.}, dielectric permittivity and ohmic conductivity by using the {\sc fem}, which were later compared to those of analytical formulae. In the simulations the concentration of the inclusion phase was constant and the shape of the inclusion was assumed to be regular polygons with $n$ sides.  The size of the polygon was used to control the {\sc fem} discretization of the computational domain.  The analytical models were based on {\sc ema} and {\sc rad}, and two different procedures were used to estimate the effective properties. It was found that the procedure based on Gauss' law and the total losses in the medium was not that succesfull as the other one based on the current and phase shift between the applied voltage and current. In order to find the polygon to mimic a disk, the finite-size scaling behavior was introduced by considering the number of sides, $n$, in the regular polygons; in reality, a regular polygon with $n$ sides becomes a disk as $n\rightarrow\infty$. It was found that for $n>15$, there was no significant change in the effective electrical quantities of mixture for low concentrations, $q\le0.3$. Consequently, when the effective quantities of mixture with an regular decagon inclusion was compated with those obtained from the analytical formulae for a similar mixture with a disk inclusion, the percentage error between quantities yield less than $0.1\ \%$. 

\section*{Acknowledgments}
Suggestions of Dr. Steven Boggs is acknowledged. Dr. Emre Tuncer is thanked for his fruitful comments. 


\end{document}